\documentclass[aps,reprint,showpacs]{revtex4-1}

\usepackage{amssymb}
\usepackage{amsmath}
\usepackage{graphicx}
\usepackage{epstopdf}

\usepackage{rotating}
\usepackage{soul}

\begin{document}


\title{Enhancement of pinning properties of superconducting thin films by graded pinning landscapes}

\author{M. Motta}
\author{F. Colauto}
\affiliation{Departamento de F\'{\i}sica, Universidade Federal de S\~{a}o Carlos, 13565-905 S\~{a}o Carlos, SP, Brazil}

\author{W. A. Ortiz}

\affiliation{Departamento de F\'{\i}sica, Universidade Federal de S\~{a}o Carlos, 13565-905 S\~{a}o Carlos, SP, Brazil} \affiliation{Centre for Advanced Study, Norwegian Academy of Science and Letters, NO-0271 Oslo, Norway}

\author{J. Fritzsche}
\affiliation{Department of Applied Physics, Chalmers University of Technology, S-412 96 G\"{o}teborg, Sweden.}

\author{J. Cuppens}
\author{W. Gillijns}
\author{V. V. Moshchalkov}

\affiliation{INPAC -- Institute for Nanoscale Physics and Chemistry, Nanoscale Superconductivity\\ and Magnetism Group, K.U.Leuven, Celestijnenlaan 200D, B--3001 Leuven, Belgium}

\author{T. H. Johansen}

\affiliation{Centre for Advanced Study, Norwegian Academy of Science and Letters, NO-0271 Oslo, Norway}\affiliation{Department of Physics, University of Oslo, POB 1048, Blindern, 0316 Oslo, Norway}

\author{A. Sanchez}

\affiliation{Departament de F\'{\i}sica, Universitat Aut\`onoma de Barcelona, 08193 Bellaterra, Barcelona, Spain}

\author{A. V. Silhanek}

\affiliation{D\'epartement de Physique, Universit\'e de Li\`ege, B-4000 Sart Tilman, Belgium}


\date{\today}
\begin{abstract}

A graded distribution of pinning centers (antidots) in superconducting MoGe thin films has been investigated by magnetization and magneto-optical imaging. The pinning landscape has maximum density at the border, decreasing progressively towards the center. At high temperatures and low fields, where this landscape mimics the vortex distribution predicted by the Bean model, an increase of the critical current is observed. At low temperatures and fields, the superconducting performance of the non-uniform sample is also improved due to suppression of thermomagnetic avalanches. These findings emphasize the relevance of non-uniform pinning landscapes, so far experimentally unexplored, on the enhancement of pinning efficiency.

\end{abstract}


\maketitle


In the mixed state of type-II superconductors, magnetic flux is admitted into the sample in the form of quantized vortices~\cite{Abrikosov57}. When superconducting currents are present, vortices undergo a viscous motion, generating a local temperature rise. This unwanted consequence of vortex motion should be prevented, in order to avoid the weakening of the superconducting properties of the material, what would constitute a threat to its potential use in real applications. For this reason, the task of understanding vortex dynamics in the presence of pinning centers (PCs) has maintained, throughout the years, its status of a timely and relevant research problem~\cite{Campbell&Evetts72,Blatter&al94} for fundamental science and applications~\cite{LargeScale74,Meingast&Larbalestier89,Baert95,Martin97,Dam&al99,Kang06,Foltyn07,Gutierrez07,Velez08,Atlas10,maiorov}. A natural strategy in this attempt to anchor vortices is to spread, at random, small clusters of normal material, an approach adopted since long for superconducting alloys~\cite{Campbell&Evetts72,Coote72,Raffy72} and reproduced more recently for high-temperature superconductors~\cite{Chong97,Zhao01,Dou02}. Other commonly employed methods are the placement of arrays of magnetic dots at the surface of superconducting films~\cite{Otani93,Martin97,Velez08}; and the creation of lattices of antidots (ADs) in films~\cite{Fiory78,Baert95,Moshchalkov96,Moshchalkov98,Welp02,Welp05,Motta11}.

It has been demonstrated~\cite{for instance1} that the insertion of arrays of ADs in a superconducting film leads, at high temperatures, to an increase of the critical current, via enhancement of the pinning efficiency. Unfortunately, at low temperatures, such PCs facilitate the proliferation of flux channeling~\cite{Motta11,Silhanek04,Menghini05} leading to unwanted instabilities of thermomagnetic origin~\cite{Denisov06,Yurchenko07} which render the superconductor impractical.

One major issue regarding the effectiveness of vortex anchorage is to adapt the PC landscape  in order to match the actual distribution of vortices. In this regard, a considerable effort has been done to investigate the case where artificial pinning sites reproduce the regular (periodic) vortex distribution, typically obtained under field cooling conditions. However, in order to create a distribution of PCs compatible with zero-field cooling conditions, one should then distribute them with a density gradient, decreasing from the edges toward the center of the sample, as expected for the vortex distribution of a partially penetrated sample in the mixed state according to the well established Bean model. Such a inhomogeneous distribution, should provide nearly optimum pinning only for a finite interval of values of the magnetic field and temperature, within which local matching conditions between the density of PCs and the density of vortices,  would be achieved. Determining the optimum gradient parameters is a non trivial problem needing to extend the standard Bean model, which assumes a constant critical current density $J_c$, in order to account for position dependent $J_c$($r$) and also to take into account the spatial dependence of demagnetizing fields.  However, even a non-optimized version of a gradient density of ADs, decreasing from the edges to the center, constitute a seemingly promising alternative way of using arrays of PCs in order to increase the critical current of the specimen.

\begin{figure}[b!]
\centering
\includegraphics[width=8.5cm]{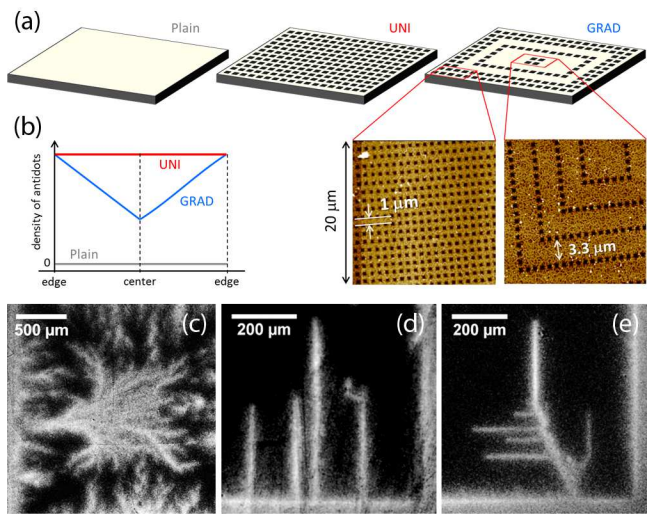}
\caption{(Color online) (a) Scheme of the density of ADs for samples  Plain, UNI, and GRAD (not to scale); (b) left panel: average areal density of
antidots as a function of position for the three cases of pinning distribution considered in this work; (b) right panels: atomic force microscopy image
showing the distribution of antidots at the border and at the center of the GRAD sample;(c) Magneto-optical image, taken at $T$ = 2.8 K and $H$ = 2.2 Oe, for a
plain sample (see Ref. ~\cite{sister sample}); (d) same as (c) for sample UNI, at $T$ = 4 K and $H$ = 1.2 Oe; (e) sample GRAD at $T$ = 4.75 K and $H$ = 1.3 Oe.}
\end{figure}

In this Letter we demonstrate that the insertion of an array of ADs with a spatially decreasing density of pinning sites (the gradient sample) promotes an increase of the critical current which is even better than the enhancement obtained when a homogeneous array of ADs (the uniform sample) is used. Furthermore, although flux avalanches on the gradient sample are induced - as expected - by the presence of the array of pinning sites, this effect is comparatively less important than for the uniform sample. This is evidenced by the substantially smaller thermomagnetic instability region on the magnetic phase diagram for the gradient sample compared to the corresponding region for the uniform sample. Thus, the presence of a grid of ADs with density gradient not only generates the desired increase in critical current, but also promotes an additional protection against the early occurrence of flux avalanches, as compared to the case of a film with a uniform distribution of ADs.

The samples investigated consist of amorphous Mo$_{79}$Ge$_{21}$ (a-Mo$_{79}$Ge$_{21}$) thin films with thickness $d=25$ nm, deposited by pulsed laser deposition on top of a Si/SiO$_2$ substrate. The pinning centers on both the gradient and the uniform sample consist of square holes of $0.5$ $\mu$m side, prepared by standard electron beam lithography. The lattice parameter of the uniform film (UNI) is 1 $\mu$m and the lattice symmetry is square. At the edges of the gradient sample (GRAD) the separation between the centers of neighboring holes is also 1 $\mu$m, and its density varies inwards (but not laterally) with a constant increase on the row separation of 10 nm/row. Figure 1(a) and Figure 1(b) left panel sketch the scheme of the spatial dependence of the density of ADs on the three films studied here. The right panels of Figure 1(b) show a zoomed up view obtained via atomic force microscopy of the edges and the center of sample GRAD, illustrating its density gradient. A third film (Plain), without ADs, was used as a reference sample.

The samples have a lithographically defined square shape with lateral dimensions of 1x1 mm$^2$ and the critical temperatures, $T_c$, are 6.73 K, 6.65 K and 7.10 K, for samples Plain, UNI and
GRAD, respectively. Using the expression for the upper critical field in the dirty limit~\cite{Schmidt97}, the zero-temperature superconducting coherence length and penetration depth were estimated for the Plain sample to be $\xi_{GL}(0) = 5$ nm and $\lambda_{GL}(0) = 517$ nm, respectively. Dc magnetization measurements were carried out in a commercial Quantum Design MPMS instrument. The magneto-optical technique employed for imaging the flux penetration morphology relies on the occurrence of the Faraday effect~\cite{Helseth01} on a magnetic indicator film placed on top of the superconducting specimen. The indicators used in the present work are Bi-substituted yttrium iron garnet films (Bi:YIG) with in-plane magnetization.

As already mentioned, within a limited interval of values of the applied field and temperature, flux avalanches originating from thermomagnetic
instabilities are likely to develop in thin films of a variety of superconducting materials~\cite{flux avalanches}. To our knowledge, however, amorphous
films of MoGe were not reported thus far as members of this list. The lower panels of Figure 1 are representative examples of such avalanches in three MoGe
films: panel 1(c) show freely expanding dendritic structures, typical of plain films~\cite{sister sample}, whereas panels 1(d) and 1(e) exhibit straight
tracks, guided by the rows of ADs~\cite{Motta11,Vlasko-Vlasov00}, in the specimens UNI and GRAD, respectively.

\begin{figure}[t!]
\begin{center}
\includegraphics[scale=1]{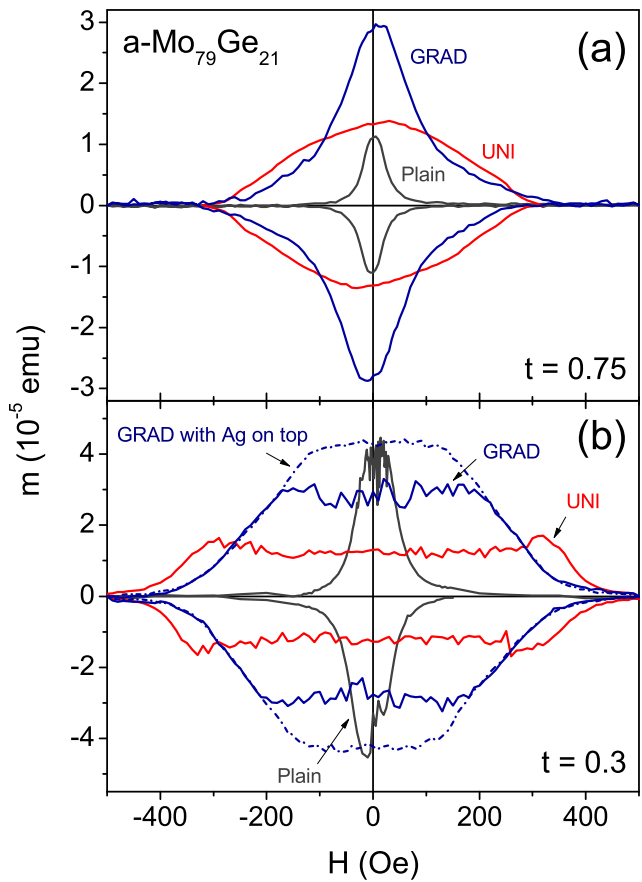}
\caption{(Color online) (a) Dc magnetization versus magnetic field taken at the reduced temperature $t = 0.75$ for samples Plain, UNI and GRAD. (b) Magnetization loops for the 3 samples in the avalanche region ($t = 0.3$); a fourth loop is also shown for sample GRAD covered with a thick disk of Ag, an artifact employed to substantially suppress flux avalanches (see text).}
\end{center}
\label{fig2}
\end{figure}

Figure 2 depicts the effects on the critical current of a film of a-Mo$_{79}$Ge$_{21}$, caused by the insertion of different arrays of ADs. Panel (a) comprises magnetization loops for the three samples studied, taken at the reduced temperature $t = T/T_c = 0.75$. At such temperatures, the loops for the samples with ADs are clearly wider and taller than that for the Plain film, confirming the enhanced pinning capability of the patterned samples, which implies larger critical currents for those specimens. It is particularly interesting to notice that sample UNI performs somewhat better than sample GRAD at magnetic field values for which the profile of penetrated flux in both films is nearly constant, i.e., $H\gg H_p$~\cite{Hp}. This behavior can be ascribed to the fact that sample UNI has more ADs than sample GRAD, so that its pinning capability is better at larger fields. At fields below 100 Oe, however, the magnetic response of sample GRAD represents an enormous enhancement of the critical current, as compared with the uniform sample. At fields around zero, its increased pinning capability leads roughly to a factor of 2 on the critical current. We have obtained similar results in Pb samples with much smaller $\lambda/\xi$ ratio (not shown). The fact that no much difference is seen in the magnetization loop for increasing and decreasing field, indicates that the graded distribution is equally efficient as a vortex dam preventing vortex entrance and exit and leading to a huge flux trapping at zero field. It is worth mentioning here that similar results were recently predicted theoretically from simulations of the critical current of superconducting specimens with somewhat more complicated landscapes of graded pinning centers~\cite{Misko&Nori12,Reichhardt12}.

The lower panel (b) in Figure 2 shows a similar set of hysteresis loops at the reduced temperature $t = T/T_c = 0.30$. The noisy response at the central portion of the loops is the typical signature~\cite{noisy loops} of the flux avalanches mentioned above. Alternative means to suppress such flux bursts have been already discussed in the literature~\cite{magnetic breaking,Colauto10}, the simplest of which we have applied here by placing a metallic disk (Ag) on top of sample GRAD, one manages to inhibit, via magnetic breaking~\cite{Colauto10}, the thermomagnetic instabilities that trigger avalanches. The dot-dashed curve in Figure 2(b) shows the hysteresis loop of sample GRAD with the Ag disk, indicating that flux avalanches are mostly suppressed in this configuration. One can thus conclude that the strategy of patterning a superconducting film with a gradient distribution of ADs is clearly more efficient in increasing the critical current than using a uniformly distributed array of ADs.

\begin{figure}[b!]
\begin{center}
\includegraphics[scale=1]{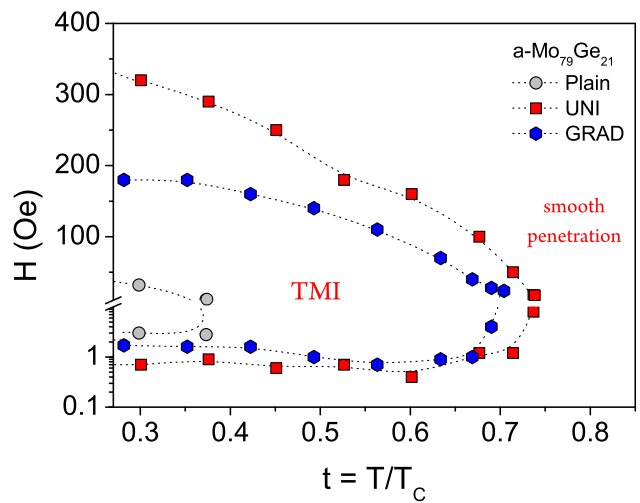}
\caption{(Color online) Boundaries of the instability region of the studied a-MoGe thin films. Notice the logarithmic scale on the lower portion of the vertical axis.}
\end{center}
\label{fig3}
\end{figure}

By repeating experiments as those depicted in Figure 2 for different values of the temperature, one can construct a $HT$-diagram containing the boundaries of the instability region for the samples studied. These frontiers are shown in Figure 3, from which one can clearly see that the inclusion of ADs enlarges the instability region, as compared to the Plain film. The region for sample GRAD is, however, substantially smaller than for sample UNI (notice the logarithmic scale on the lower portion of the vertical axis). Since these avalanches can be efficiently suppressed by depositing a metallic layer on top of the film of interest, one can fully appreciate the reach of the present results, from which we conclude that the strategy of inserting a graded pinning landscape represents an increased protection of the film against flux avalanches, a substantial advantage in terms of practical use of superconducting films in the presence of perpendicular magnetic fields. At the moment it remains unclear to us whether this improved performance of the sample GRAD arises from the lack of perfect periodicity in the antidot lattice or from the gradient itself.

In summary, we have demonstrated that a graded pinning landscape introduced in a superconducting film of a-Mo$_{79}$Ge$_{21}$ increases the critical current, as compared to a uniform distribution of ADs. In addition, flux avalanche activity, typically induced by the presence of arrays of ADs, is less prejudicial for the sample with gradient distribution of antidots than for the uniformly distributed pinning centers. This work focus on a particular gradient geometry following a very smooth linear decrease of areal density of pinning centers from the border of the sample towards its center. Further investigations in other gradient geometries including non-linear areal-density dependence or inverted gradients will be necessary to identify the main mechanisms and geometrical parameters responsible for the striking improvement of the pinning properties of this sort of pinning landscape.

This work was partially supported by the Brazilian funding agencies FAPESP and CNPq, the Fonds de la Recherche Scientifique - FNRS, the Methusalem Funding of the Flemish Government, the Fund for Scientific Research-Flanders (FWO-Vlaanderen), the program for scientific cooperation F.R.S.-FNRS-CNPq and Spanish projects CSD2007-00041 and MAT2012-35370.


\begin{thebibliography}{99}
\bibitem{sister sample}
Flux avalanches in superconducting films develop only below an upper threshold temperature, which is characteristic of each sample. For the sample named Plain in this work, this threshold temperature is  2.6 K, as determined by magnetic measurements, a value that we cannot  achieve in the cold-finger type cryostat of the magneto-optical imaging setup. We present instead an avalanche observed on a sister plain film of larger dimensions (2.5 mm x 2.5 mm), at T = 2.8 K and H = 2.2 Oe.
\bibitem{Hp}
The characteristic field $H_p$, which serves as a reference for flux penetration through the center of the film, is 2 Oe for all samples studied at 5 K. It might be roughly determined as the field at which the virgin curve is matched by a subsequent measurement with increasing positive fields (not shown in Fig.2 to avoid excessive information in the drawings).
\end{thebibliography}
\end{document}